\documentclass[preprint,aps,prc,tightenlines,nofootinbib,showpacs]{revtex4}
\usepackage{bm}
\begin{document}
\input{psfig}
\input{epsf}
\def\Im{\mbox{\sl Im\ }}
\def\pd{\partial}
\def\oln{\overline}
\def\olft{\overleftarrow}
\def\ds{\displaystyle}
\def\bgreek#1{\mbox{\boldmath $#1$ \unboldmath}}
\def\sla#1{\slash \hspace{-2.5mm} #1}
\newcommand{\bra}{\langle}
\newcommand{\ket}{\rangle}
\newcommand{\vep}{\varepsilon}
\newcommand{\met}{{\mbox{\scriptsize met}}}
\newcommand{\lab}{{\mbox{\scriptsize lab}}}
\newcommand{\cm}{{\mbox{\scriptsize cm}}}
\newcommand{\mcal}{\mathcal}
\newcommand{\Del}{$\Delta$}
\newcommand{\g}{{\rm g}}
\long\def\Omit#1{}
\long\def\omit#1{\small #1}
\def\beq{\begin{equation}}
\def\eeq{\end{equation} }
\def\bea{\begin{eqnarray}}
\def\eea{\end{eqnarray}}
\def\eqref#1{Eq.~(\ref{eq:#1})}
\def\eqlab#1{\label{eq:#1}}
\def\figref#1{Fig.~\ref{fig:#1}}
\def\figlab#1{\label{fig:#1}}
\def\tabref#1{Table \ref{tab:#1}}
\def\tablab#1{\label{tab:#1}}
\def\secref#1{Section~\ref{sec:#1}}
\def\appref#1{Appendix~\ref{sec:#1}}
\def\seclab#1{\label{sec:#1}}
\def\VYP#1#2#3{{\bf #1}, #3 (#2)}  
\def\NP#1#2#3{Nucl.~Phys.~\VYP{#1}{#2}{#3}}
\def\NPA#1#2#3{Nucl.~Phys.~A~\VYP{#1}{#2}{#3}}
\def\NPB#1#2#3{Nucl.~Phys.~B~\VYP{#1}{#2}{#3}}
\def\PL#1#2#3{Phys.~Lett.~\VYP{#1}{#2}{#3}}
\def\PLB#1#2#3{Phys.~Lett.~B~\VYP{#1}{#2}{#3}}
\def\PR#1#2#3{Phys.~Rev.~\VYP{#1}{#2}{#3}}
\def\PRC#1#2#3{Phys.~Rev.~C~\VYP{#1}{#2}{#3}}
\def\PRD#1#2#3{Phys.~Rev.~D~\VYP{#1}{#2}{#3}}
\def\PRL#1#2#3{Phys.~Rev.~Lett.~\VYP{#1}{#2}{#3}}
\def\FBS#1#2#3{Few-Body~Sys.~\VYP{#1}{#2}{#3}}
\def\AP#1#2#3{Ann.~of Phys.~\VYP{#1}{#2}{#3}}
\def\ZP#1#2#3{Z.\ Phys.\  \VYP{#1}{#2}{#3}}
\def\ZPA#1#2#3{Z.\ Phys.\ A\VYP{#1}{#2}{#3}}
\def\half{\mbox{\small{$\frac{1}{2}$}}}
\def\quarter{\mbox{\small{$\frac{1}{4}$}}}
\def\nn{\nonumber}
\newlength{\PicSize}
\newlength{\FormulaWidth}
\newlength{\DiagramWidth}
\newcommand{\vslash}[1]{#1 \hspace{-0.42 em} /}
\newcommand{\qslash}[1]{#1 \hspace{-0.46 em} /}
\def\her{\marginpar{$\Longleftarrow$}}
\def\bel{\marginpar{$\Downarrow$}}
\def\abo{\marginpar{$\Uparrow$}}



\title{P- and T-violating $\pi N N$ form factor} 

\author{S.~Kondratyuk}
\affiliation{Department of Physics and Astronomy, University of Manitoba,
Winnipeg, MB, Canada R3T 2N2}
\author{P.~G.~Blunden}
\affiliation{Department of Physics and Astronomy, University of Manitoba,
Winnipeg, MB, Canada R3T 2N2}

\date{\today}

\begin{abstract}
The form factor of the parity and time-reversal violating (PTV) pion-nucleon interaction is calculated from one-loop vertex diagrams. The degrees of freedom included in the
effective lagrangian are nucleons, pions, $\eta$, $\rho$ and $\omega$ mesons.
We show that by studying the form factor one can constrain the 
PTV meson-nucleon coupling constants.
We evaluate the mean square radius associated with the PTV $\pi N N$ vertex.
Using the mean square radius, we estimate the effect of the 
PTV $\pi N N$ vertex on the neutron electric dipole moment, and find a very small
correction. We also extract the renormalisation group $\beta$ function 
and use it to discuss evolution of the PTV $\pi N N$ coupling constant
beyond the hadronic mass scale. 
\end{abstract}

\pacs{11.30.Er, 13.75.Gx, 24.80.+y}
\maketitle


\section{Introduction} \seclab{intro}
 
The parity (P) and time-reversal (T) nonconserving pion-nucleon coupling
is a theoretical ingredient in many calculations of electric dipole moments (EDMs)
of various systems, such as the neutron, deuteron,
various atoms and molecules~\cite{Khr97}. The continuing 
interest in the problem of EDMs, and its relevance
to possible extensions of the Standard Model~\cite{Pos05}, is a major motivation for
a deeper understanding of the P- and T-violating (PTV) $\pi N N$ interaction.
Usually in calculations one uses a point-like PTV $\pi N N$ interaction, 
neglecting a form factor associated with effects of nonlocality.

In this paper we calculate such a form factor 
from meson loop corrections to a bare PTV $\pi N N$ vertex. 
The loop integrals are evaluated in a relativistic field theoretical approach based on
an effective lagrangian comprising the nucleon, pion, $\rho$,
$\omega$ and $\eta$ mesons. The lagrangian consist of a strong interaction piece involving
in principle well-known coupling constants, and a
PTV piece involving unknown coupling constants.
Our aim is to study various features of the calculated PTV $\pi N N$ form factor -- for example, its
dependence on the square of the momentum transferred by the pion to the nucleon, the associated mean square radius, etc.  
Using renormalisation group methods in a one-loop approximation,
we also consider possible scenarios of the
high-energy evolution of the PTV $\pi N N$ constant. The latter problem in particular 
is motivated by the puzzle of the extremely small value of the charge and parity (CP) violating $\theta$ term in the QCD lagrangian~\cite{Pos05}.
Throughout the paper, we do not
assign specific numerical values to the unknown PTV meson-nucleon coupling constants. 
We do, however, make general assumptions as to the relations among them (e.~g.~choosing the constants to be of comparable magnitude or
assuming some of them to be suppressed) and analyse consequences of these assumptions for the
PTV $\pi N N$ form factor, the mean square radius and the high-energy scale evolution
of the PTV coupling constant. 

Our results could be used to  
study effects of the PTV $\pi N N$ form factor in
calculations of electric dipole moments of the nucleon. 
We will make a tentative estimate indicating that the PTV $\pi N N$ form factor 
introduces a very small correction to a leading chiral contribution to the neutron EDM. 
To our knowledge, 
a measurement of a PTV $\pi N N$ form factor has not been attempted so far.
We will show that such a measurement could, in particular, put some constraints among
the PTV meson-nucleon coupling constants.  
Since at present such experimental applications are rather speculative,
in this paper we will focus on the PTV form factor itself,
leaving specific calculations of related measurable PTV amplitudes for future research.

\section{Effective lagrangian}\seclab{lagr}

The interaction among hadrons in our model is described by the sum of the 
strong lagrangian ${\mathcal{L}_S}$ and 
the P-,T-violating lagrangian ${\mathcal{L}_{\qslash{P}\qslash{T}}}$~\cite{Liu04}:
\beq
{\mathcal{L}}={\mathcal{L}_S}+ {\mathcal{L}_{\qslash{P}\qslash{T}}},
\eqlab{lagr}
\eeq
where
\bea
{\mathcal{L}_S} &=& \frac{g_\pi}{2 M} \oln{N} \gamma_5 \gamma_\mu 
{\bm{\tau}}\cdot (\pd^\mu {\bm{\pi}}) N 
+\frac{g_\eta}{2 M} \oln{N} \gamma_5 \gamma_\mu (\pd^\mu \eta) N 
+ i g_{\rho \pi} {\bm{\rho}}_\mu \cdot [{\bm{\pi}} \times \pd^\mu {\bm{\pi}} ] \nn \\
&+& g_\rho \oln{N} \left\{ (\gamma^\mu - \frac{\kappa_\rho}{2 M} \sigma^{\mu \nu} \pd_\nu )
{\bm{\tau}} \cdot {\bm{\rho}}_\mu \right\} N 
+ g_\omega \oln{N} \left\{ (\gamma^\mu - \frac{\kappa_\omega}{2 M} \sigma^{\mu \nu} 
\pd_\nu ) \omega_\mu \right\} N , \eqlab{lagr_s}
\eea   
\bea
{\mathcal{L}_{\qslash{P}\qslash{T}}} &=& \oln{N} \left\{ \oln{g}_\pi^{(0)} 
{\bm{\tau}}\cdot{\bm{\pi}} + \oln{g}_\pi^{(1)} \pi^0 +
\oln{g}_\pi^{(2)} \tau_3 \pi^0 \right\} N  +
\oln{N} \left\{ \oln{g}_\eta^{(0)} \eta + \oln{g}_\eta^{(1)} \tau_3 \eta \right\} N \nn \\
&+& \frac{i}{2 M} \oln{N} \left\{ \sigma^{\mu \nu} \gamma_5 \pd_\nu
\left[ \oln{g}_\rho^{(0)} {\bm{\tau}}\cdot{\bm{\rho}}_\mu + \oln{g}_\rho^{(1)} \rho_\mu^0 +
\oln{g}_\rho^{(2)} \tau_3 \rho_\mu^0 \right] 
\right\} N \nn \\
&+& \frac{i}{2 M} \oln{N} \left\{ \sigma^{\mu \nu} \gamma_5 \pd_\nu
\left[ \oln{g}_\omega^{(0)} \omega_\mu + \oln{g}_\omega^{(1)} \tau_3 \omega_\mu \right] 
\right\} N \eqlab{lagr_pt}
\eea 
(our constants $\oln{g}_{\pi,\rho}^{(0)}$ and $\oln{g}_{\pi,\rho}^{(2)}$ 
are equal to $\oln{g}_{\pi,\rho}^{(0)} - \oln{g}_{\pi,\rho}^{(2)}$ and
$3 \oln{g}_{\pi,\rho}^{(2)}$, respectively, as defined in Ref.~\cite{Liu04}).  
The nucleon field is denoted as $N$; the pion, $\rho$, $\omega$ and $\eta$ meson
fields as $\pi^i$, $\rho^i_\mu$, $\omega_\mu$ and $\eta$, respectively, with the
lower indices denoting Lorentz vector components and the upper indices isovector
components; the bold font indicates isovectors and $\tau_i$ are the 
Pauli isospin matrices.\footnote{All other conventions and definitions used throughout this paper follow Ref.~\cite{Bjo65}.}
The hadron masses are: $M=0.938$~GeV for the nucleon, $m=0.138$~GeV for the pion, 
$m_\rho=0.77$~GeV, $m_\omega=0.78$~GeV, $m_\eta=0.55$~GeV for the other mesons.
The coupling constants in the strong lagrangian 
${\mathcal{L}}$ are chosen as in Ref.~\cite{Liu04}: $g_\pi=13.07$, $g_\eta=2.24$,
$g_\rho=2.75$, $\kappa_\rho=3.7$, $g_\omega=8.25$, $\kappa_\omega=-0.12$ and
$g_{\rho \pi}=6.07$ (from the width of the $\rho$ resonance~\cite{PDG06}).

Retaining all isospin structures in the
PTV $\pi N N$ (and other meson-nucleon) vertices would involve many 
unknown coupling constants, which would make results of loop calculations based on
the lagrangian~\eqref{lagr_pt} difficult to interpret.
Therefore, to keep our analysis tractable, in the following we will consider 
only the $\sim \oln{g}_\pi^{(0)}$ isospin structure in the PTV $\pi N N$ vertex,
which involves a charged pion and therefore appears in the calculations of the neutron 
EDM~\cite{Cre79,Liu04}.  
We will use the simplified notation for the renormalised (physical)
coupling constants: $c_\pi=\oln{g}_\pi^{(0)}$,
$c_\rho=\oln{g}_\rho^{(0)}$, $c_\omega=\oln{g}_\omega^{(0)}$, 
$c_\eta=\oln{g}_\eta^{(0)}$. Note that the PTV $\pi N N$ coupling constant in
the lagrangian~\eqref{lagr_pt} is a bare constant $c_\pi^B$ which will serve to cancel
infinities of the loop integrals in the course of renormalisation
described in the next section. 

We use the pseudovector ({\em pv}) strong $\pi N N$ vertex in~\eqref{lagr_s}.
The pseudoscalar ({\em ps}) structure 
$ i g_\pi \oln{N} \gamma_5 {\bm{\tau}}\cdot {\bm{\pi}} N$
is also used in the literature on 
effective lagrangian calculations of effects of parity and time reversal 
violation~\cite{Cre79,Liu04}. The {\em pv} structure of the $\pi N N$ vertex was used
in the recent chiral calculation~\cite{Hoc05} of the nucleon electric dipole form factor.  
While the {\em ps} and {\em pv} couplings are equivalent on shell, the pseudovector is more consistent with chiral constraints, which motivates our choice.
In the course of the discussion below
we will specify important differences and similarities between the results obtained using the {\em pv} and the {\em ps} strong $\pi N N$ vertices.

\section{One-loop vertex diagrams generating the form factor}\seclab{loops}

We calculate the vertex loop diagrams depicted in~\figref{loops}, 
for the space-like pion momenta, $q^2 \equiv -Q^2 \le 0$, and with
both nucleons on-shell.
\begin{figure}[!htb]
\centerline{{\epsfxsize 9.5cm \epsffile[30 180 580 610]{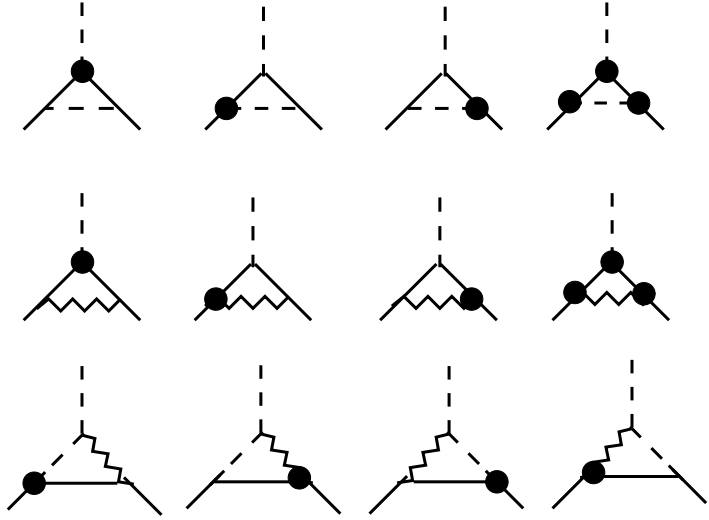}}}
\caption[f1]{The loop diagrams included in the calculation of the P- and T-violating
$\pi N N$ form factor. The solid and dashed lines denote nucleons and pions, respectively.
In addition, in the upper row, the internal dashed lines are $\pi$ or $\eta$ mesons, and in the middle row, the zigzag lines are $\rho$ or $\omega$ mesons. The blobs represent 
the PTV interactions from~\eqref{lagr_pt}, and the other vertices denote
the strong interactions from~\eqref{lagr_s}.   
\figlab{loops}}
\end{figure}

Before renormalisation, the PTV vertex can be written as $c_\pi^B + \Gamma^B(q^2)$, which  
is the sum of the tree-level contribution $c_\pi^B$ from the
lagrangian~\eqref{lagr_pt} and the bare (unrenormalised) one-loop corrections $\Gamma^B(q^2)$.
To render the calculated loop integrals $\Gamma^B(q^2)$ ultraviolet convergent,
we apply the modified minimal subtraction procedure in conjunction with dimensional regularisation~\cite{tHo72,Wei95}.
This procedure amounts to setting the constant
$\Delta$ (defined in~\appref{pv_fun}), which is divergent in the limit $D \rightarrow 4$, to zero in the unrenormalised loops $\Gamma^B(q^2)$.
In this way we obtain ultraviolet and infrared convergent loop integrals which do not depend on the regularisation parameters (see~\appref{pv_fun}): the dimension $D$ (which 
after renormalisation can be set to the physical value $D=4$) and the interim cutoff 
$\Lambda$ (which may be thought of as a scale much larger than the included hadronic masses). 
The renormalisation can be interpreted as an absorption of the infinities of the unrenormalised loops $\Gamma^B(q^2)$
into the bare coupling constant $c_\pi^B$, such that the full vertex is rewritten as the sum of a renormalised coupling constant $c_\pi$ and a renormalised loop contribution 
$\Gamma(q^2)$, both of them free of infinities. The PTV $\pi N N$ form factor 
$F_{\qslash{P}\qslash{T}}(Q^2)$ is then defined by 
\beq
c_\pi + \Gamma(q^2) = c_\pi F_{\qslash{P}\qslash{T}}(Q^2) \,.
\eqlab{ff_def}
\eeq

The sum of the renormalised loop diagrams with an intermediate pion 
reads 
\beq
\Gamma_\pi(q^2) = \frac{c_\pi g_\pi^2}{16 \pi^2} 
\left\{V_\pi(q^2) - \frac{1}{4}\right \} 
+ \frac{c_\pi^3}{16 \pi^2} 
\left\{ V_\pi(q^2) + 4 M^2 C_0[M^2,q^2,M^2;m^2,M^2,M^2]  \right\} \,,
\eqlab{v_pi}
\eeq
where 
\bea
V_\pi(q^2) &=& B_0[0;M^2,M^2] + \frac{2 q^2 B_0[M^2;m^2,M^2]}{4M^2-q^2}-
\frac{(4M^2+q^2) B_0[q^2;M^2,M^2]}{4M^2-q^2} \nn \\
&-& \frac{m^2 (4M^2+q^2) C_0[M^2,q^2,M^2;m^2,M^2,M^2]}{4M^2-q^2} \,,
\eqlab{aux_pi}
\eea
with $B_0$ and $C_0$ the two- and three-point Passarino-Veltman 
functions~\cite{tHo79}, given in~\appref{pv_fun}.
It can be easily verified that any dependence on the regularisation parameters 
$\Delta$ and $\Lambda$ cancels out in~\eqref{v_pi}
(as it should), thus proving the ultraviolet and infrared convergence of the
renormalised loop. 
Using formulae from~\appref{pv_fun} and retaining only the leading orders of
$m/M$ and $(m/M)\,\mbox{ln}(m/M)$ in the $B_0$ functions, 
\eqref{v_pi} can be rewritten in a more explicit form
\bea
\Gamma_\pi(q^2) & \approx & -\frac{c_\pi g_\pi^2}{64 \pi^2} +
\frac{c_\pi}{16 \pi^2 (4 M^2-q^2)} \left\{ (g_\pi^2+c_\pi^2) 
\left[ 4 q^2 \left( 1-\frac{m}{M}\mbox{arctg}\frac{M}{m} \right) 
-2 q^2 \frac{m^2}{M^2} \mbox{ln} \frac{m}{M} \right. \right. \nn \\
&+& \left. (4 M^2+q^2) \int_0^1\!dx \, \mbox{ln}\left(1+\frac{q^2 x (x-1)}{M^2} \right) 
 \right] \nn \\
&+& \!\! \left[ 4M^2(4M^2-q^2)c_\pi^2-m^2(4M^2+q^2)(g_\pi^2+c_\pi^2) \right] \!
C_0[M^2,q^2,M^2;m^2,M^2,M^2] \Bigg\} ,
\eqlab{v_pi_explic}
\eea
with $C_0[M^2,q^2,M^2;m^2,M^2,M^2]$ given by~\eqref{c0_spec}. (The $q^2$-independent additive constant is unimportant here since in presenting our results we will normalise 
$F_{\qslash{P}\qslash{T}}(Q^2=0)=1$.) 
The vertex loops with the
$\rho$, $\omega$ and $\eta$ mesons have a structure similar to 
that of the pion in~\eqref{v_pi}; their contributions are given in~\appref{lps_heavy}. 

Examples of the calculated PTV $\pi N N$ form factor 
$F_{\qslash{P}\qslash{T}}(Q^2)$  
are shown in~\figref{mes_eff}, where we display separately the
contributions from the vertex loops from~\figref{loops} involving 
$\pi$, $\rho$, $\omega$ and $\eta$ mesons.  
The form factor depends on a particular relation among the generally unknown
PTV meson-nucleon coupling constants $c_\pi$, $c_\eta$, $c_\rho$ and $c_\omega$. 
This is illustrated in the four panels of~\figref{mes_eff}, where we used
PTV constants of comparable size with various sign combinations.
\begin{figure}[!htb]
\centerline{{\epsfxsize 14cm \epsffile[30 150 580 720]{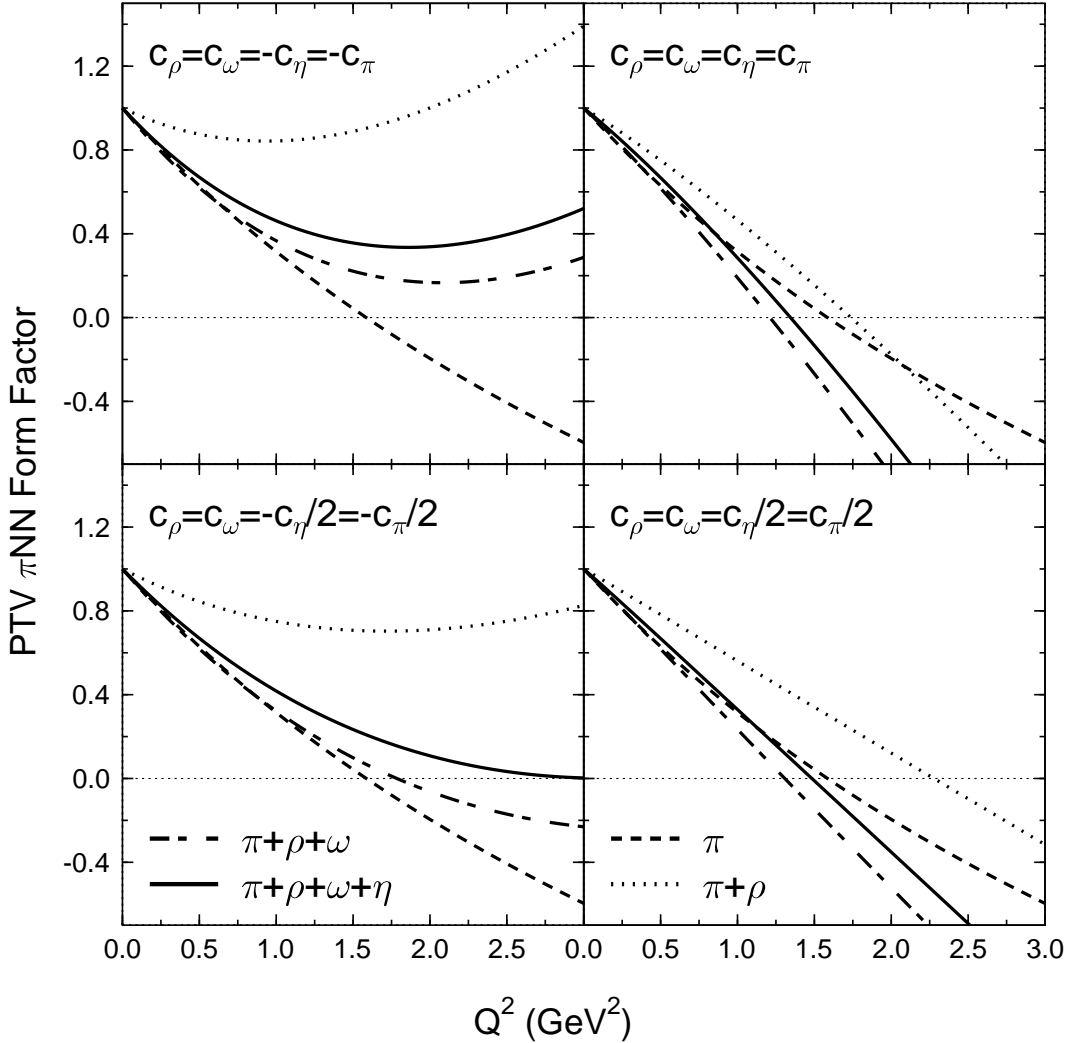}}}
\caption[f2]{Contributions of the meson loops to the
P- and T-violating $\pi N N$ form factor $F_{\qslash{P}\qslash{T}}(Q^2)$, 
defined in~\eqref{ff_def}, using the indicated
relations among the PTV meson-nucleon coupling constants.
The total form factor is the sum of all meson contributions and is
given by the solid curves.
\figlab{mes_eff}}
\end{figure}
The results in~\figref{mes_eff} were obtained using
the pseudovector ({\em pv}) structure of the strong $\pi N N$ vertex, 
as given in~\eqref{lagr_s}. If instead we use the pseudoscalar ({\em ps})
$\pi N N$ vertex 
to calculate the diagrams in~\figref{loops}, then only the loops
containing the vector mesons $\rho$ and $\omega$ would be different from the {\em pv} case. 
As a result, there is an approximate relation between the PTV $\pi N N$ form factors obtained using the {\em ps} and the {\em pv} $\pi NN$ vertices: the form factor obtained using
the {\em ps} $\pi N N$ vertex and with a certain relation among the PTV constants 
$c_\pi$, $c_\eta$, $c_\rho$ and $c_\omega$ is approximately equal to that obtained using the {\em pv} vertex, but with the signs of $c_\rho$ and $c_\omega$ changed. 
For example, the lower left panel in~\figref{mes_eff} shows either the form factor calculated using the {\em pv} vertex and with
$c_\rho=c_\omega=-0.5 c_\eta =-0.5 c_\pi$ or (approximately) the form factor obtained using the {\em ps} vertex, but with $c_\rho=c_\omega=0.5 c_\eta =0.5 c_\pi$.

We can see from~\figref{mes_eff} that the contributions of the vector mesons $\rho$ and 
$\omega$ to the PTV $\pi N N$ form factor are
larger when $c_\rho$ and $c_\omega$ have opposite signs compared to $c_\pi$ and $c_\eta$
of the pseudoscalar mesons. Furthermore, for any combination of the
PTV coupling constants, the $\rho$ and $\omega$ contributions have comparable magnitudes but opposite signs. The $\eta$ meson gives the smallest contribution.
If the PTV coupling constants $c_\rho$, $c_\omega$ and $c_\eta$ were all much smaller than 
$c_\pi$, then the resulting form factor would be approximated by
the pion loop contribution alone, given by the dashed line in~\figref{mes_eff}.
Currently there is no strong reason to assume that $c_\pi$ is dominant, or adopt any other particular relation among the PTV coupling constants (various arguments on this issue have been presented in Refs.~\cite{Gud93,Liu04}).
Our results in~\figref{mes_eff} indicate that a measurement of the PTV $\pi N N$ form factor
could be very useful for establishing such a relation.
\figref{mes_eff} also shows that for some
relations among the coupling constants, the form factor may have a zero at
$Q^2$ of the order $\sim M^2$.
Although at present we cannot make a specific statement about the significance of 
such a zero, it might be important in some applications of the form factor.
In the next two sections we will analyse in more detail the PTV $\pi N N$ form factor in 
the low- and high-energy regions.

\section{Mean square radius and the effect on the neutron EDM}\seclab{rad}

We define the mean square radius associated with the PTV $\pi N N$ interaction by analogy with the standard definition used for electromagnetic form factors:
\beq
\left< r^2_{\qslash{P}\qslash{T}} \right> = 
-6 \left( \frac{\pd F_{\qslash{P}\qslash{T}}}{\pd Q^2} \right)_{Q^2=0}.
\eqlab{rad_gen}
\eeq

The contributions to the radius from the loops involving particular mesons
will be denoted by the corresponding subscripts: 
\beq
\left< r^2_{\qslash{P}\qslash{T}} \right>_{\pi} \approx 
\left( 0.19 -0.00062 \, c_\pi^2 \right) \,\mbox{fm}^2 \,,
\eqlab{rad_pi}
\eeq
\beq
\left< r^2_{\qslash{P}\qslash{T}} \right>_{\rho} \approx 
\left( -0.092 \pm 0.015 \, {c_{\rho} \over c_\pi} +0.00035 \, c_\rho^2 \right) \,\mbox{fm}^2 \,,
\eqlab{rad_rho}
\eeq
\beq
\left< r^2_{\qslash{P}\qslash{T}} \right>_{\omega} \approx 
\left( 0.093 \mp 0.032 \, {c_\omega \over c_\pi} - 0.00036 \, c_\omega^2 \right) \,\mbox{fm}^2 \,,
\eqlab{rad_om}
\eeq
\beq
\left< r^2_{\qslash{P}\qslash{T}} \right>_{\eta} \approx 
\left( 0.0011 - 0.026 \, {c_\eta \over c_\pi} - 0.000018 \, c_\eta^2 \right) \,\mbox{fm}^2 \,,
\eqlab{rad_eta}
\eeq
where the upper (lower) sign in Eqs.~(\ref{eq:rad_rho}) and (\ref{eq:rad_om})
corresponds to the use of the {\em pv} ({\em ps}) structure of the strong $\pi N N$ vertex.
The leading term of
the pion loop contribution can be written analytically in a compact form 
\beq
\left< r^2_{\qslash{P}\qslash{T}} \right>_{\pi} = 
\frac{5 g_\pi^2}{16 \pi^2 M^2 }  + {\mathcal{O}}(c_\pi^2, m^2/M^4) =
\frac{5 g_A^2}{16 \pi^2 F_{\pi}^2 } + {\mathcal{O}}(c_\pi^2, m^2/M^4) 
\approx 0.22 \, \mbox{fm}^2 \,,
\eqlab{rad_pi_chir}
\eeq
where we have used the Goldberger-Treiman relation 
$g_\pi F_\pi = g_A M$ (with $F_\pi=93$ MeV being the pion decay constant and $g_A=1.267$
the axial coupling constant), valid in leading chiral order.
Neglecting the (presumably suppressed) quadratic terms in 
Eqs.~(\ref{eq:rad_pi}-\ref{eq:rad_eta}) and
adding up the leading terms, we obtain the
total mean square radius associated with the PTV $\pi N N$ form factor:
\beq
\left< r^2_{\qslash{P}\qslash{T}} \right> = 
\left< r^2_{\qslash{P}\qslash{T}} \right>_{\pi}+
\left< r^2_{\qslash{P}\qslash{T}} \right>_{\rho}+
\left< r^2_{\qslash{P}\qslash{T}} \right>_{\omega}+
\left< r^2_{\qslash{P}\qslash{T}} \right>_{\eta} \approx
\left( 0.19 \pm 0.015  {c_\rho \over c_\pi} \mp 0.032  {c_\omega \over c_\pi}
- 0.026 {c_\eta \over c_\pi} \right) \mbox{fm}^2.
\eqlab{rad_sum}
\eeq
This shows, in particular, that if $|c_{\rho,\omega,\eta}| {\,\slash} {\!\!\!\!\gg} |c_\pi|$, 
then the dominant contribution to $\left< r^2_{\qslash{P}\qslash{T}} \right>$ comes almost entirely from the pion loops, in which case the explicit formula~\eqref{rad_pi_chir} can be used as a good approximation to the full result~\eqref{rad_sum}: 
$\left< r^2_{\qslash{P}\qslash{T}} \right> \approx
\left< r^2_{\qslash{P}\qslash{T}} \right>_\pi$.

We can use the mean square radius to estimate the effect of the 
PTV $\pi N N$ form factor on the neutron EDM $D_n$. 
First we will cite here some results on $D_n$ obtained in approaches similar to ours, 
where we will use the recent reviews~\cite{Pos05,Tim06} and references therein. 
The upper bound from experiments with ultra-cold neutrons is 
$D_n^{EXP} \le 7.5 \times 10^{-26} e \cdot \mbox{cm}$,
where $e$ denotes the elementary electric charge.
In the Standard Model calculations, $D_n$ appears as a strongly suppressed 
perturbation effect: $D_n^{SM} \approx 10^{-32} e \cdot \mbox{cm}$.
If a discrepancy between the Standard Model prediction and experiment is eventually established,
it may indicate a P- and T-violation, or 
other fundamental extensions of the Standard Model. These possibilities are explored in many 
calculations wherein one uses EDM measurements to constrain various PTV parameters, 
e.~g.~the QCD $\theta$-term, CP-violating quark coupling constants, etc. 
Examples of such model calculations, most relevant in the context of the present paper,
include the following. 
The MIT quark bag model~\cite{Bal79} yields the neutron dipole moment
$D_n^{QM} \approx 8.2 \times 10^{-16}\, \theta \,e \cdot \mbox{cm}$. 
The estimate from the leading chiral logarithm~\cite{Cre79} is
$D_n^{\chi Log} \approx 3.6 \times 10^{-16}\, \theta \, e \cdot \mbox{cm}$, while the extended
$SU(3)$ chiral perturbation theory calculation~\cite{Pic91} gives to first order
$D_n^{\chi PT} \approx (3.3 \pm 1.8) \times 10^{-16}\, \theta \, e \cdot \mbox{cm}$.
In the effective lagrangian model~\cite{Liu04} $D_n$ is evaluated in terms of the PTV 
meson-nucleon coupling constants given in~\eqref{lagr_pt}:
$D_n^{EL} \approx \left[ 0.14 \, \oln{g}_{\pi}^{(0)} - 
0.02 \, (\oln{g}_{\rho}^{(0)} - \oln{g}_{\rho}^{(1)} + \oln{g}_{\rho}^{(2)} ) +
6 \times 10^{-3} \oln{g}_{\omega}^{(0)} \right] e \cdot \mbox{cm}$. 
Another approach~\cite{Pos00}, 
based on QCD sum rules, predicts
$D_n^{SR} \approx 1.2 \times 10^{-16} \, \theta \, e \cdot \mbox{cm}$. 

The aforementioned leading chiral contribution $D_n^{\chi Log}$
is obtained~\cite{Cre79,Pic91,Hoc05} from the one-pion loop correction to the PTV 
$\gamma NN$ vertex where the photon couples to the intermediate pion:
\beq
D_n^{\chi Log} = \frac{e c_\pi g_\pi}{4 \pi^2 M} \, \mbox{ln}\frac{M}{m} \,, 
\eqlab{edm_0}
\eeq
with effects of the PTV $\pi N N$ form factor being neglected in~\eqref{edm_0}.
We assume that the loop integral for the neutron EDM is dominated by the region of
very small four-momenta squared $Q^2$ of the intermediate pions.
In this region the form factor 
$F_{\qslash{P}\qslash{T}}(Q^2) \approx 1 - \left< r^2_{\qslash{P}\qslash{T}} \right> Q^2/6$. 
If we set $Q^2 \approx m^2$ in this expansion and 
substitute it into the EDM pion-loop correction in place of the 
PTV $\pi N N$ form factor, then the $Q^2$ dependence of the form factor does not have to be integrated. Thus we obtain an estimate for the neutron EDM
\beq
D_n^{FF} = D_n^{\chi Log} \, \left( 1+\delta^{FF}_{\qslash{P}\qslash{T}} \right) \,,
\eqlab{edm_eff}
\eeq
with the correction due to the PTV $\pi N N$ form factor
\beq
\delta^{FF}_{\qslash{P}\qslash{T}} 
\approx -\frac{m^2}{6} \left< r^2_{\qslash{P}\qslash{T}} \right>  
\approx -\frac{5 g_\pi^2 m^2}{96 \pi^2 M^2} + {\mathcal{O}}(c_\pi^2, m^4/M^4) 
\approx -2 \% \,,
\eqlab{del_ff}
\eeq
where we have used~\eqref{rad_pi_chir}.
We should point out that in view of the preceding simplifying assumptions, this is a rather crude estimate which may
give only an approximate magnitude of the effect.
A more detailed treatment must involve a careful integration of the 
EDM vertex loop including the
PTV $\pi N N$ form factor (which in turn depends on the unknown PTV
meson-nucleon coupling constants, as discussed in~\secref{loops}).
Such a calculation falls outside the scope of this paper.

\section{High-energy scale evolution of the PTV $\pi N N$ coupling 
constant from the Callan-Symanzik equation}\seclab{beta}

The Callan-Symanzik (or renormalisation group) equation~\cite{Cal70,Wei95,Pes95} describes the evolution of a renormalised coupling constant $c_\pi(\mu)$, defined at a typical mass scale $\sqrt{-q^2} \sim \mu$, where, ideally, all the masses in the problem are negligible in comparison with
$\mu$:
\beq
\mu \frac{d c_\pi(\mu)}{d \mu} = \beta(c_\pi) \,.
\eqlab{callsym_eq}
\eeq
In our case it should be sufficient to take 
$\mu$ to be much larger than the nucleon mass $M$.
The evolution is determined by the renormalisation-group 
$\beta$ function $\beta(c_\pi) \equiv \beta_\pi$ which we will calculate in this section.
To keep the discussion transparent, we will consider the scale evolution of 
the PTV $\pi N N$ coupling constant in isolation, extracting the $\beta$
function from the one-pion loops only and ignoring the evolution
of the other coupling constants.
Although these simplifications make the 
renormalisation group considerations somewhat formal, the results of this section
should provide a useful approximation of the true behaviour of $c_\pi(\mu)$.
For definiteness, throughout this section we assume that $c_\pi(\mu) > 0$.
This choice of the sign does not restrict 
the renormalisation group considerations since, as~\eqref{callsym_eq} shows,
$\beta(-c_\pi)=-\beta(c_\pi)$. In general, one can consider
$|c_\pi(\mu)|$ without any change in the following discussion.

We will calculate the $\beta$ functions for the two structures of the strong (non-PTV) 
$\pi N N$ vertex: the {\em pv} structure, given in~\eqref{lagr_s}, 
and the {\em ps} structure. As described in~\secref{loops}, the renormalised
pion loops for the PTV form factor are the same using either 
of the two structures. However, the $\beta$ functions are different in these two cases, as  will be shown below.
We use dimensional regularisation to calculate the loop integrals at
a space-time dimension $D$ and apply the modified minimal
subtraction procedure. 
In this scheme, the $\beta$ function at one-loop level  
can be extracted as the coefficient at the term 
$1/(4-D)-\gamma_E/2+\mbox{ln}(4 \pi)/2 \equiv \Delta/2$ which enters 
into the bare coupling constant in order to cancel the UV divergences
of the loop integrals~\cite{Wei95,Pes95,tHo73}.
The $\Delta$ term appears in our calculation only through the functions $B_0$
in the {\em unrenormalised} loop integrals, see \eqref{b0_explic}. 
Thus we obtain:
\beq
\beta_\pi^{(pv)}=\frac{3 g_\pi^2}{16 \pi^2} c_\pi - \frac{1}{8 \pi^2} c_\pi^3 \,,
\eqlab{beta_pi_pv}
\eeq
using the {\em pv} strong $\pi N N$ vertices in the loops, and
\beq
\beta_\pi^{(ps)}=-\frac{g_\pi^2}{8 \pi^2} c_\pi - \frac{1}{8 \pi^2} c_\pi^3 \,,
\eqlab{beta_pi_ps}
\eeq
using the {\em ps} vertices.

Integrating the Callan-Symanzik equation (\ref{eq:callsym_eq}) between mass scales 
$\mu_1$ and $\mu_2$, with the $\beta$ 
functions~\eqref{beta_pi_pv} and~\eqref{beta_pi_ps}, we obtain the solution
\beq
\left( \frac{\mu_2}{\mu_1} \right)^A = \frac{c_\pi(\mu_2)}{c_\pi(\mu_1)}
\sqrt{\frac{1 + {B \over A} c_\pi^2(\mu_1)}{1 + {B \over A} c_\pi^2(\mu_2)}} \,,
\eqlab{callsym_sol}
\eeq
where
\beq
A=A^{(pv)}=\frac{3 g_\pi^2}{16 \pi^2} \approx 3.25\,,\;\; 
B=-\frac{1}{8 \pi^2} \approx -0.013\,,
\eqlab{ab_pv}
\eeq
and
\beq
A=A^{(ps)}=-\frac{g_\pi^2}{8 \pi^2} \approx -2.16\,,\;\; 
B=-\frac{1}{8 \pi^2} \approx -0.013\,,
\eqlab{ab_ps}
\eeq
using the {\em pv} and {\em ps} strong $\pi N N$ vertices, respectively.

If we work at moderate mass scales $\mu$, where $|B c_\pi^2(\mu) /A| \ll 1$,
we can retain only the leading order in the $\beta$ function,
\beq
\beta_\pi \approx A \,c_\pi \,,
\eqlab{beta_low}
\eeq
for which an approximate solution of~\eqref{callsym_eq} can be written
\beq
c_\pi(\mu) \approx c_\pi(\mu_0) \left(\frac{\mu}{\mu_0}\right)^A \,,
\eqlab{callsym_low}
\eeq
where $\mu_0$ is some ``initial" renormalisation point (one may choose it
to be of the order of the nucleon mass, $\mu_0 \sim M$, assuming that 
$c_\pi(\mu_0 \sim M)$ has a magnitude relevant to 
EDM measurements or to other PTV experiments).

Consider first the calculation using the {\em pv} strong $\pi N N$ vertex in the loops. In this case $A=A^{(pv)} > 0$, see~\eqref{ab_pv}, 
hence the approximate 
solution~\eqref{callsym_low} describes a growth of $c_\pi(\mu)$ with $\mu$. 
For a sufficiently large scale $\mu$ the value of $c_\pi(\mu)$ could become so big that
the $\beta$ function~\eqref{beta_pi_pv} could vanish. Then the 
approximate solution~\eqref{callsym_low} would no longer be valid. 
Instead,~\eqref{callsym_eq} shows that
$c_\pi(\mu \rightarrow \infty)$ would tend to a fixed point $c_\pi^*$
determined from the condition $\beta_\pi^{(pv)}(c_\pi^*)=0$. In the vicinity of the
fixed point 
\beq
\beta_\pi^{(pv)} \approx a (c_\pi^* - c_\pi)\,,\;\;\;
c_\pi^*=\sqrt{\frac{3}{2}}g_\pi \approx 16 \,,\;\;
a=\frac{3 g_\pi^2}{8 \pi^2} \approx 6.5\,,
\eqlab{beta_high}
\eeq
and the solution of~\eqref{callsym_eq} describes how $c_\pi(\mu)$ approaches $c_\pi^*$:
\beq
c_\pi^* - c_\pi(\mu) \sim  \left( \frac{\mu}{\mu_0} \right)^{-a}\,.
\eqlab{callsym_high}
\eeq
Recalling that $c_\pi$ is proportional to the CP-violating $\theta$
term which could be included in a QCD lagrangian~\cite{Cre79,Pos05}, the evolution
of $c_\pi(\mu)$ in \eqref{callsym_low} is compatible with an extremely small 
$\theta \le 3 \times 10^{-10}$~\cite{Pos05,Khr97}, as extracted from EDM measurements
at low energies.
In this scenario, obtained in our model using the {\em pv} strong $\pi N N$ vertex, 
an initially small value of $c_\pi$ grows as a positive power of the mass scale $\mu$ and may eventually reach a fixed value $c_\pi^*$. Using~\eqref{callsym_low} with 
$A=A^{(pv)} \approx 3.25$, $\mu_0 \approx M$, $c_\pi(M) \le 7 \times 10^{-12}$~\cite{Khr97} 
and $c_\pi(\mu)=g_\pi/\sqrt{2}$ (the point at which $\beta_\pi^{(pv)}$ of~\eqref{beta_pi_pv} reaches its maximum and starts decreasing), one can estimate that the fixed point $c_\pi^*$ would be approached at energies above $\mu \sim 5$ TeV.

Now consider the calculation with the {\em ps} strong $\pi N N$ vertices in the loops
for the PTV form factor. 
A qualitatively different behaviour of $c_\pi(\mu)$ obtains in this case. The $\beta$ 
function~\eqref{beta_pi_ps} remains negative for all $c_\pi$, and the
general solution of the renormalisation group equation~\eqref{callsym_eq} is well approximated by~\eqref{callsym_low}. Since $A=A^{(ps)} < 0$, see~\eqref{ab_ps}, the 
solution~\eqref{callsym_low} describes a gradual decrease and eventual vanishing of 
$c_\pi(\mu)$ as $\mu \rightarrow \infty$.

Whether one of these two contrasting scenarios of the high-energy scale evolution of
$c_\pi(\mu)$ is correct should be, in principle, clarified by
experiment. 
An important caveat here is that the dynamics at high energies will certainly involve
hadron resonances, quark-gluon and other -- possibly supersymmetric --
degrees of freedom not included in our model 
(for a recent review, see~\cite{Pos05} and references therein). 
These degrees of freedom would probably influence the high-energy evolution of $c_\pi(\mu)$, 
which in turn could change the discussion presented in this section.

\section{Summary} \seclab{summ}

We have described a calculation of the form factor associated with the parity and 
time-reversal violating (PTV) pion-nucleon coupling. The form factor is
generated in our model by one-loop vertex corrections involving nucleon, 
pion, $\rho$-, $\omega$- and $\eta$-meson degrees of freedom. The interaction of these hadrons is described by
a phenomenological lagrangian consisting of a strong interaction part and a
PTV part. The latter is determined by several PTV meson-nucleon coupling constants,
all of which are in general unknown, as are their relative orders of magnitude.

One result of the present calculation
is that by measuring the PTV $\pi N N$ form factor, it should in principle be possible to constrain this set of the PTV meson-nucleon coupling constants, since different
relations among them lead to distinct $Q^2$ dependences of the form factor, 
as shown in~\figref{mes_eff}. 
The experimental feasibility of such a measurement is 
certainly a challenging problem in its own right, 
and we have left an investigation of this question outside the scope of the
present paper.
We found that the PTV $\pi N N$ form factor 
generated by the loops with the strong {\em pv} $\pi N N$ vertex is related in a simple
way to that generated by the loops with the strong {\em ps} vertex: these two form factors are converted into each other by reversing the signs of the PTV couplings of the vector mesons. 

We have analysed the low-energy characteristics of the PTV pion-nucleon interaction by
evaluating the mean square radius associated with its form factor. It turned out that,
if all the PTV coupling constants are of the same order of magnitude (or at least if
the pion-nucleon constant is not much smaller than the others), then the contributions of
the $\rho$ and $\omega$ mesons almost completely cancel each other, and that of the
$\eta$ meson is suppressed. As a result, the mean square radius can be 
well approximated by the pion-loop contribution alone, which can be given by a simple expression~\eqref{rad_pi_chir}.
This is consistent with the intuitive expectation that the pion cloud should dominate the
low-energy features of the interaction.
The simplified estimate in~\secref{rad}
suggests that the effect of the PTV $\pi N N$ form factor on the neutron EDM is 
very small. A more definite statement can be made on the basis of a detailed calculation of the loop integrals for the EDM, including the full $Q^2$-dependence of the PTV 
$\pi N N$ form factor, which was not pursued in the present paper.

It has been shown elsewhere~\cite{Cre79,Pic91} that the PTV pion-nucleon coupling constant is proportional to the CP-violating $\theta$ term which can be added to the QCD lagrangian. The current estimates 
of $\theta$ based on measurements of EDMs of the neutron and various atoms and molecules, yield an extremely small upper limit $|\theta| < 3 \times 10^{-10}$,
which presents a long-standing puzzle~\cite{Pos05}. Since the value of $\theta$ may be energy-scale dependent (for example, yielding appreciable CP-violating
effects at sufficiently high energies), we have considered in our model 
possible scenarios of the evolution of the 
PTV $\pi N N$ coupling constant $c_\pi(\mu)$ with the energy scale $\mu$.
To this end, we employed renormalisation group techniques, with several simplifying assumptions. Being used in the framework of a model, results of such an analysis are
to some extent speculative. Nevertheless, we found in~\secref{beta} an intriguing possibility that, in the calculation with the {\em pv} strong $\pi N N$
vertex, the PTV $\pi N N$ coupling constant relevant to very high energy scales may be significantly 
larger than the tiny value extracted from the low-energy EDM measurements.
In the ultraviolet limit $\mu \rightarrow \infty$, 
$c_\pi(\mu)$ may eventually reach a fixed value $c_\pi^*$,~\eqref{beta_high},
of a ``natural" size.

%

\appendix

\section{Passarino-Veltman functions}\seclab{pv_fun}

In this appendix we give the scalar Passarino-Veltman 
functions~\cite{tHo79,Anl01} 
$B_0$ and $C_0$ entering into the expressions for the loop integrals.
The functions are defined as D-dimensional integrals
\beq
B_0[p^2;m_1^2,m_2^2] \equiv \frac{(2 \pi \Lambda)^{4-D}}{i \pi^2} 
\int \! d^D q \left\{ [q^2-m_1^2 + i0] [(q+p)^2-m_2^2+i0] \right\}^{-1} \,,
\eqlab{b0_def}
\eeq
$$
C_0[p_1^2,p_2^2,(p_1+p_2)^2;m_1^2,m_2^2,m_3^2] \equiv
$$
\beq
\frac{(2 \pi \Lambda)^{4-D}}{i \pi^2}
\int \! d^D q \left\{ [q^2-m_1^2+i0][(q+p_1)^2-m_2^2+i0][(q+p_1+p_2)^2-m_3^2+i0] 
\right\}^{-1} ,
\eqlab{c0_def}
\eeq
where $\Lambda$ is an arbitrary cutoff mass introduced to keep the dimensions of $B_0$ and 
$C_0$ independent of $D$ (upon renormalisation, the loop integrals do not depend on
$\Lambda$, and the limit $D \rightarrow 4$ can also be taken without encountering divergences). Eqs.~(\ref{eq:b0_def}) and (\ref{eq:c0_def})
can be reduced to one- and two-dimensional integrals which are either evaluated numerically or expressed via logarithm and dilogarithm functions. We have
\beq
B_0[p^2;m_1^2,m_2^2] = \Delta + B_0^f[p^2;m_1^2,m_2^2] \,,
\eqlab{b0_explic}
\eeq
where  
\beq
\Delta \equiv \frac{2}{4-D} - \gamma_E + \mbox{ln}(4 \pi),\;\; 
\gamma_E \approx 0.5772 \,\,(\mbox{Euler's constant}).
\eqlab{delta_def}
\eeq
is a divergent (in the limit $D \rightarrow 4$)
constant typical for the modified minimal subtraction procedure, and 
\beq
B_0^f[p^2;m_1^2,m_2^2]= - \int_0^1 \! dx \,
\mbox{ln} \left( \frac{x^2 p^2 - x(p^2-m_2^2+m_1^2)+m_1^2-i0}{\Lambda^2} \right) + 
{\mathcal{O}}(D-4) \,,
\eqlab{b0R_explic}
\eeq
is a finite (ultraviolet convergent) one-dimensional integral.

In our loop integrals we encounter only $C_0$ functions with $p_1^2=(p_1+p_2)^2=M^2$
and $p_2^2=q^2$,
which can be reduced to convergent two-dimensional integrals 
$$
C_0[M^2,q^2,M^2;m_1^2,m_2^2,m_3^2] = 
$$
\beq - \int_0^1 \! dx \int_0^{1-x} \! dy
\left\{ M^2(x+y)^2-xy q^2 -M^2 (x+y) -x(m_1^2-m_2^2)-
y(m_1^2-m_3^2)+m_1^2 -i0 \right\}^{-1} \!\!\!.
\eqlab{c0_explic}
\eeq

The $B_0$ and $C_0$ functions from the pion loops~\eqref{v_pi} can be further simplified:
\bea
B_0^f[q^2;M^2,M^2] &=& - 2 \, \mbox{ln} \frac{M}{\Lambda} -
\int_0^1 \! dx \, \mbox{ln}\left( 1+\frac{q^2 x(x-1)}{M^2} - i0 \right) ,
\eqlab{b0_spec3}
\eea 
$$
C_0[M^2,q^2,M^2;m^2,M^2,M^2]=
$$
\beq
\frac{x_q}{M^2(1-x_q^2)} 
\left\{ \mbox{ln}(x_q) \left[2 \, \mbox{ln}(1+x_q) - \frac{1}{2}\,\mbox{ln} \,x_q - 
2 \, \mbox{ln} \frac{m}{M} \right] 
+ \frac{\pi^2}{6} + 2 \, \mbox{Li}_2(-x_q) \right\} \,,
\eqlab{c0_spec}
\eeq
where $x_q = \left[ \sqrt{(q^2-4M^2)/(q^2+i0)} -1 \right] \cdot
\left[ \sqrt{(q^2-4M^2)/(q^2+i0)} +1 \right]^{-1}$ and  
the dilogarithm (Spence) function is 
\beq
\mbox{Li}_2(x) = -\int_0^1 \! dy \, \frac{\mbox{ln}(1-xy)}{y} \;\;\; 
(\, |\mbox{arg}(1-x)| < \pi \,) \,.
\eqlab{spence}
\eeq
The following special cases are useful in deriving 
Eqs.~(\ref{eq:v_pi_explic}) and (\ref{eq:rad_pi_chir}):
\beq
B_0^f[q^2;M^2,M^2] = - 2 \, \mbox{ln} \frac{M}{\Lambda} + 
\frac{q^2}{6M^2} + {\mathcal{O}}\left(\frac{q^4}{M^4}\right) \;\;\; 
(\,|q^2| \ll M^2\,) \,,
\eqlab{b0_spec2}
\eeq
\beq
B_0^f[M^2;m^2,M^2] = - 2 \, \mbox{ln} \frac{M}{\Lambda} + 
2 \left( 1 - \frac{m}{M} \mbox{arctg} \frac{M}{m} \right) - 
\frac{m^2}{M^2} \, \mbox{ln} \frac{m}{M}  + 
{\mathcal{O}}\left(\frac{m^2}{M^2}\right) ,
\eqlab{b0_spec}
\eeq
\beq
C_0[M^2,0,M^2;m^2,M^2,M^2] = \frac{1}{M^2}\, \mbox{ln} \frac{m}{M} 
+ {\mathcal{O}}\left( \frac{m^2}{M^4} \right) .
\eqlab{c0_spec1}
\eeq

\section{Loop integrals with the $\rho$, $\omega$ and $\eta$ mesons}\seclab{lps_heavy}

In this appendix we give explicit formulae for the renormalised loop diagrams
with $\rho$, $\omega$ and $\eta$ mesons, where for brevity we will 
retain only the dominant linear terms $\sim c_{\pi,\rho,\omega,\eta}$, 
marking them with a superscript ``(1)". The contribution from the 
$\eta$ meson equals
\beq
\Gamma_\eta^{(1)}(q^2)  =  
\frac{c_\eta g_\pi g_\eta q^2}{8 \pi^2 (4 M^2 -q^2)} 
\!\left\{ V_\eta(q^2) -B_0[M^2;M^2,m_\eta^2]  \right\} 
+ \frac{c_\pi g_\eta^2}{16 \pi^2} 
\!\left\{ V_\eta(q^2) - B_0[0;M^2,M^2] + \frac{1}{4} \right\} ,
\eqlab{v_eta}
\eeq
where
\beq
V_\eta(q^2)=B_0[q^2;M^2,M^2]
+ m_\eta^2 \, C_0[M^2,q^2,M^2;m_\eta^2,M^2,M^2] \,.
\eqlab{aux_eta}
\eeq
The contributions from the $\omega$ and $\rho$ mesons read
\bea
\Gamma_\omega^{(1)}(q^2) & = & 
-\frac{c_\pi g_\omega^2}{384 M^4 \pi^2}
\left( 48 M^4 +120 \kappa_\omega M^4+
54 \kappa_\omega^2 M^4 +12 \kappa_\omega^2 M^2 m_\omega^2 + 8 M^2 q^2 \right. \nn \\
&+& \left. 24 \kappa_\omega M^2 q^2 + 10 \kappa_\omega^2 M^2 q^2 
+ \kappa_\omega^2 m_\omega^2 q^2 \right) \nn \\
& - & \frac{g_\omega}{128 M^4 \pi^2}
\left\{ c_\pi g_\omega \left( 16 M^4 +48 \kappa_\omega M^4 +
20 \kappa_\omega^2 M^4 +2 \kappa_\omega^2 M^2 m_\omega^2 + 4 M^2 q^2 \right. \right. \nn \\
&-& \left. \left. 3 \kappa_\omega^2 M^2 q^2 + 
\kappa_\omega^2 m_\omega^2 q^2 \right) + 
c_\omega g_\pi q^2 \left( 4 M^2 +  6 \kappa_\omega M^2
- \kappa_\omega m_\omega^2 \right) \right\} B_0[0;M^2,M^2] \nn \\
& \!\!\!\!\!\!\!\!\!\!\!\!\!\!\! - & \!\!\!\!\!\!\!\!\!\! 
\frac{g_\omega q^2}{128 M^4 \pi^2 (4 M^2-q^2)} 
\left\{ c_\pi g_\omega \left( 4 M^2 + \kappa_\omega^2 m_\omega^2 \right) +
c_\omega g_\pi \left( 4 M^2 - \kappa_\omega m_\omega^2 \right) \right\}
B_0[M^2;M^2,m_\omega^2] \nn \\
& + & \frac{g_\omega}{128 M^2 \pi^2 (4 M^2 - q^2)}
\left\{ c_\pi g_\omega \left( 64 M^4 + 192 \kappa_\omega M^4 + 
80 \kappa_\omega^2 M^4 \right. \right. \nn \\
&+& \left. \left. 8 \kappa_\omega^2 M^2 m_\omega^2 -
48 \kappa_\omega M^2 q^2 - 32 \kappa_\omega^2 M^2 q^2 +
2 \kappa_\omega^2 m_\omega^2 q^2 + 3 \kappa_\omega^2 q^4 \right) \right. \nn \\
&+& \left. 2 c_\omega g_\pi q^2 \left( 8 M^2 + 12 \kappa_\omega M^2 -
2 \kappa_\omega m_\omega^2 - 3 \kappa_\omega q^2 \right) \right\}
B_0[q^2;M^2,M^2] \nn \\
& + & \frac{g_\omega}{64 M^2 \pi^2 (4 M^2 -q^2)}
\left\{ c_\pi g_\omega \left( 64 M^6 + 32 M^4 m_\omega^2 +
96 \kappa_\omega M^4 m_\omega^2 \right. \right. \nn \\ 
&+& 32 \kappa_\omega^2 M^4 m_\omega^2 
+ 4 \kappa_\omega^2 M^2 m_\omega^4 - 48 M^4 q^2 -
24 \kappa_\omega M^2 m_\omega^2 q^2 - 16 \kappa_\omega^2 M^2 m_\omega^2 q^2 \nn \\
&+& \left. \kappa_\omega^2 m_\omega^4 q^2 + 8 M^2 q^4 +
2 \kappa_\omega^2 m_\omega^2 q^4 \right) \nn \\
&+& \left. \!\! 2 c_\omega g_\pi m_\omega^2 q^2 \left( 4 M^2 + 8 \kappa_\omega M^2 -
\kappa_\omega m_\omega^2 - 2 \kappa_\omega q^2 \right) \right\} \!
C_0[M^2,q^2,M^2;m_\omega^2,M^2,M^2] ,  
\eqlab{v_om}
\eea
\bea
\Gamma_{\rho}^{(1)}(q^2) & = &  
-\left( \Gamma_{\omega}^{(1)}(q^2) \right)_{m_\omega \rightarrow m_\rho, 
c_\omega \rightarrow c_\rho, g_\omega \rightarrow g_\rho, \kappa_\omega \rightarrow 
\kappa_\rho}  \nn \\
&\!\!\!\!\!\!\!\!\!\!\!\!\!\!\! + &\!\!\!\!\!\!\!\!\!\! 
\frac{g_{\rho \pi} c_\rho g_\pi (m^2 -M^2)}{32 M^2 \pi^2} 
\left\{ 1 + B_0[0;m^2,M^2] \right\} -
\frac{g_{\rho \pi} c_\pi g_\rho (\kappa_\rho m^2 -4 M^2 -\kappa_\rho m_\rho^2)}
{16 M^2 \pi^2} B_0[0;m^2,m_\rho^2] \nn \\
&+& \frac{g_{\rho \pi} c_\rho g_\pi}{32 \pi^2} 
\left\{ \frac{m^2}{M^2} B_0[0;m^2,m^2] - B_0[0;M^2,M^2] \right\} + 
\frac{g_{\rho \pi} c_\rho g_\pi (M^2-m_\rho^2)}{16 M^2 \pi^2} B_0[0;M^2,m_\rho^2] \nn \\
&-& \frac{g_{\rho \pi}}{32 M^2 \pi^2 (4 M^2 -q^2)} 
\left\{ c_\pi g_\rho q^2 \left( \kappa_\rho m^2 - 8 M^2 - 4 \kappa_\rho M^2 -
\kappa_\rho m_\rho^2 \right) \right. \nn \\
&+& \left. 2 c_\rho g_\pi \left( 2 m^2 M^2 + 2  M^2 m_\rho^2 - m^2 q^2 - 
2 M^2 q^2 \right) \right\} B_0[M^2;m^2,M^2] \nn \\
&+&\frac{g_{\rho \pi}}{32 M^2 \pi^2 (4M^2-q^2)}
\left\{  c_\pi g_\rho q^2 \left( 8 M^2 + 4 \kappa_\rho M^2 + \kappa_\rho m_\rho^2 -
\kappa_\rho m^2 \right) \right. \nn \\
&+& \left. 2 c_\rho g_\pi \left( 2 m^2 M^2 + 2 M^2 m_\rho^2 + 2 M^2 q^2 - 
m_\rho^2 q^2 \right) \right\} B_0[M^2;M^2,m_\rho^2] \nn \\
&-&\frac{g_{\rho \pi}}{4 \pi^2 (4 M^2 - q^2 )}
\left\{ c_\pi g_\rho \left( 4 M^2 + \kappa_\rho m_\rho^2 + q^2 + 
\kappa_\rho q^2 - \kappa_\rho m^2 \right) \right. \nn \\
&+& \left. c_\rho g_\pi \left( m^2 - m_\rho^2 +q^2 \right) \right\}
B_0[q^2;m^2,m_\rho^2] \nn \\
&+&\frac{g_{\rho \pi}}{8 \pi^2 (4 M^2-q^2)}
\left\{ c_\pi g_\rho \left( 8 M^2 m_\rho^2 - 8 m^2 M^2 + \kappa_\rho m_\rho^4 -
\kappa_\rho m^4 + 4 m^2 q^2 \right. \right. \nn \\
&+& \left. 2 \kappa_\rho m^2 q^2 - 8 M^2 q^2 - \kappa_\rho q^4 \right) \nn \\
&+& \left. c_\rho g_\pi \left( m^2 + m_\rho^2 - q^2 \right) 
\left( m^2 - m_\rho^2 + q^2 \right) \right\} C_0[M^2,q^2,M^2;M^2,m^2,m_\rho^2] \,,
\eqlab{v_pirho}
\eea
with $B_0$ and $C_0$ defined in \appref{pv_fun}. To facilitate 
the calculation of the loop integrals, we used
the computer package ``FeynCalc"~\cite{Mer91}.


\end{document}